\documentclass[a4paper,10pt]{article}
\usepackage[utf8]{inputenc}
\usepackage[toc,page]{appendix}
\usepackage{epigraph}
\usepackage{amsmath}
\usepackage{amssymb}
\usepackage{amsthm}
\usepackage{mathtools}
\usepackage{dsfont}
\usepackage{bbm}
\usepackage{array}
\usepackage{url}
\usepackage{lipsum}
\usepackage{graphicx}
\usepackage{physics}
\usepackage{caption}
\usepackage{subcaption}

\newtheorem{proposition}{Proposition}

\newtheorem{remark}{Remark}
\newtheorem{example}{Example}

\usepackage[a4paper, total={6in, 10in}]{geometry}

\DeclareMathOperator*{\argmax}{arg\,max}

\usepackage[colorlinks, linkcolor={blue!50!blue}, citecolor={red}]{hyperref}
\usepackage{todonotes}
\newcommand\1{{\mathds 1}}

\def\C{{\mathbb C}}

\def\bbI{{\mathbb I}}

\def\N{{\mathbb N}}

\def\R{{\mathbb R}}
\def\RR{{\mathbb R}}
\def\SS{{\mathbb S}}
\def\TT{{\mathbb T}}

\def\Z{{\mathbb Z}}
\def\ZZ{{\mathbb Z}}

\def\rd{{\mathrm{d}}}
\def\re{{\mathrm{e}}}
\def\ri{{\mathrm{i}}}

\def\cA{{\mathcal A}}
\def\cB{{\mathcal B}}

\def\cH{{\mathcal H}}

\def\cR{{\mathcal R}}

\newcommand{\ie}{{\em i.e.}\ }
\newcommand{\Vect}{{\rm Vect} \, }
\newcommand{\Ran}{{\rm Ran} \, }
\renewcommand{\Re}{{\rm Re} \, }
\newcommand{\U}{{\rm U}}
\newcommand{\SU}{{\rm SU}}
\newcommand{\obs}{{\rm obs}}
\newcommand{\bu}{u}
\newcommand{\homotopy}[1]{#1} 

\title{Numerical construction of Wannier functions through homotopy}
\author{D. Gontier, A. Levitt \& S. Siraj-dine}

\begin{document}
\maketitle

\abstract

We provide a mathematically proven, simple and efficient
algorithm to build localised Wannier functions, with the only
requirement that the system has vanishing Chern numbers. Our algorithm is able to build localised Wannier for topological
insulators such as the Kane-Mele model. It is based on an
explicit and constructive proof of homotopies for the
unitary group. We provide numerical tests validating the methods
for several systems, including the Kane-Mele model.

\section{Introduction}

The occupied states of a periodic model of independent electrons are
described by Bloch waves, which are (delocalised) modulated plane waves. Wannier functions are localised combinations of
Bloch waves that span the occupied space. Due to the gauge
freedom for the Bloch waves, Wannier functions are non-unique, and
their localisation properties depend strongly on the choice of gauge.
A specific gauge choice ensuring localisation was made in
\cite{marzari1997maximally}. These maximally-localised Wannier
functions (MLWF) are useful as a conceptual tool, to interpret bonding
and polarisation in crystals~\cite{king1993theory}, as well as a
numerical tool, to construct effective tight-binding
models~\cite{marzari1997maximally} and compute exact exchange
terms~\cite{wu2009order}. Methods to construct these
MLWFs enable their routine use as a
post-processing of density functional theory computations in solids.
We refer to~\cite{marzari2012maximally} for a review on applications.

The existence of localised Wannier functions for insulators is not
guaranteed. Through the Bloch transform, it is equivalent to the
following problem: given a smooth family of rank-$N$ projectors $P(k)$
defined on the $d$-dimensional torus $\mathds{T}^{d}$, can we find a
smooth Bloch frame representing the range of $P(k)$, \ie a set of $N$
orthogonal smooth functions
$\bu(k) := \left(u_1(k), \ldots, u_N(k) \right)$ on $\mathds{T}^{d}$ such
that $\Ran P(k) = \Vect \bu (k)$ for all $k \in \mathds{T}^{d}$. If it is
indeed possible, then the inverse Bloch transform of $\bu(\cdot, x)$
yields localised Wannier functions.

In dimensions $d \geq 2$, such problems involve a competition between
local smoothness and global periodicity. This is because the space $\Ran P(k)$ might twist
with $k$, analogous to the twisting of the tangent space of a Möbius
strip. Accordingly, there might be topological obstructions to
finding such a Bloch frame. These obstructions are characterised by Chern numbers (one number
in $d=2$, three numbers in $d=3$). In dimension $2$ and $3$, it is
possible to construct localised Wannier functions if and only if the
Chern numbers vanish \cite{Brouder2007exp,panati2007triviality}. In systems with
time-reversal symmetry, one has the additional property that
\begin{center}
    (TRS) \qquad $P(-k) = \theta P(k) \theta^{-1}$, where $\theta$ is an anti-unitary operator.
\end{center}
This implies that all Chern numbers vanish, and it is therefore
possible to construct Wannier functions for such systems. By
constrast, Chern insulators (a simple model of which is the Haldane
model \cite{Haldane}), characterised by a broken time-reversal
symmetry and non-trivial Chern invariants, can not support localised
Wannier functions.

A further distinction appears depending on the type of time-reversal
symmetry: bosonic (BTRS, occuring
for instance in electrons whose spin degrees of freedom are neglected)
or fermionic (FTRS, when spin-orbit coupling
is present). Mathematically, these different types are characterised
by $\theta^{2} = 1$ (BTRS) or $\theta^{2} = -1$ (FTRS). In the FTRS
case, but not in the BTRS case, a further topological obstruction
appears when trying to find Wannier functions respecting the natural
symmetry of the problem \cite{Fiorenza2016}. In $d=2$, there are two
classes of systems: those for which one can find localised symmetric
Wannier functions and those for which this is not possible. This is
reflected by the $\mathbb{Z}_{2}$-valued Fu-Kane-Mele invariant~\cite{fu2007topological}.
Physically, this appears as symmetry-protected edge states.





In the common case of BTRS (when spin-orbit coupling is ignored and
electrons pairs can be considered as spinless particles), several
algorithms exist to compute localised Wannier functions. The most
popular one was introduced by Marzari and
Vanderbilt~\cite{marzari1997maximally}. This optimises the locality of
Wannier functions, starting from an initial guess. This algorithm is
able to yield exponentially localised Wannier
functions~\cite{panati2013bloch}, but depends strongly on the choice
of the initial guess. Recent advances, based on the use of
matrix logarithms \cite{cances2017robust}, selected columns of the
density matrix (SCDM \cite{damle2015compressed,damle2017scdm}) or an
extended set of projections \cite{mustafa2015automated}, provide ways
to automatically construct initial projections, without any specific
physical input.

However, in the topologically non-trivial FTRS case, such as the
Kane-Mele model of topological insulators, substantial difficulties
appear. Since no symmetric Bloch frame can exist, algorithms that do
not explicitly break this symmetry fail. This means that the initial
guess for the method of \cite{marzari1997maximally} has to break this
symmetry manually, which often proves challenging in practice. In the
method of \cite{cances2017robust}, this manifests as a crossing of
eigenvalues, making the logarithm ill-defined (see Appendix of
\cite{cances2017robust}). In the SCDM method, this appears as a loss
of rank, unless a system-specific choice of columns is imposed
\cite{lin_private}.

The goal of this paper is to provide an automatic method that
constructs localised Wannier functions even in the FTRS case. Our
method is based on a standard reduction of the problem of finding
Wannier functions to that of finding homotopies in the unitary group $\U(N)$.
This problem was solved using matrix logarithms in
\cite{cances2017robust}, and using a multi-step logarithm based on a
perturbation argument in \cite{cornean2016construction,
  cornean2017construction}. In this paper, we instead use a iterative
method where the columns of the unitaries are contracted one by one. This method, which is natural and robust, implements an
idea hinted at, but not detailed,
in~\cite[p.81]{fiorenza2016construction}. Unlike the similar method of
\cite{cornean2016construction, cornean2017construction}, it does not
exploit the eigenstructure, which proves unstable in practice.

We emphasise that methods to construct Wannier functions specifically for the
case of $\mathbb{Z}_{2}$ insulators were proposed in
\cite{soluyanov2011wannier}, \cite{soluyanov2012smooth},
\cite{winkler2016smooth} and \cite{mustafa2016automated}. These
methods however require model-specific information, while our method
is completely automatic.

The paper is organised as follows. We present in Section
\ref{sec:wannier_to_bloch} the definition of Wannier functions, and
the equivalence between localised Wannier functions and smooth Bloch
frames. Then, we recall in Section \ref{sec:bloch_to_homotopies} the
standard reduction from the problem of finding smooth Bloch frames to
that of finding homotopies of unitary matrices. We explain in Section
\ref{sec:homotopies_and_logs} the difficulties of this problem and our
solution, which we illustrate numerically in Section
\ref{sec:num_res}.

\section{From Wannier functions to Bloch frames}
\label{sec:wannier_to_bloch}

\subsection{The Schrödinger equation in crystals}

The goal of Wannier functions is to represent the subspace of occupied
orbitals of a $d$-dimensional periodic Hamiltonian $H$ with localised
functions. More specifically, we consider a $d$-dimensional periodic
crystal described by a lattice $\mathcal{R}\sim (2\pi\ZZ)^d$. We
denote by $\mathcal{A}\sim \RR^d /(2\pi\ZZ)^d$ its unit cell, by
$\mathcal{R}^*\sim \ZZ^d$ its reciprocal lattice, and by
$\mathcal{B}\sim \RR^d/\ZZ^d$ the reciprocal unit cell, also called
the Brillouin zone. The behaviour of independent electrons (or
electrons in mean-field approaches such as Kohn-Sham density
functional theory) is described by the linear Schr\"odinger operator
$H$, given by
\[H = -\frac{1}{2}\Delta + V, \quad \text{acting on} \quad L^2(\R^d,
  \mathbb{C}),
\] 
where $V$ is a (sufficiently well-behaved) $\mathcal{R}$-periodic potential modelling the external
(mean-field) potential. Here, we dropped the spin variable for simplicity, as it plays no role in the argument.

As $H$ commutes with $\cR$-translations, it follows from Bloch-Floquet theory~\cite{ReedSimon4} that $H$ is described with its fibers $H(k)$, which, in our case, are operators acting on $\cR$-periodic functions, and given by
\begin{align*}
H(k) = \frac{1}{2}(-\ri\nabla + {k})^2 + V \quad \text{acting on} \quad L^2(\cA, \C).
\end{align*}
For all $k \in \cB$ and $K \in \cR'$, the operators $H(k)$ and $H(k+K)$ are unitarily equivalent:
\begin{equation} \label{eq:unitary_equivalent}
  H(k+K) = \tau_K H(k) \tau_K^* \quad \text{with} \quad \tau_K (u)(x) := \re^{- \ri K \cdot x}u(x).
\end{equation}
The operators $H_k$ have a compact resolvent, with eigenvalues
$\varepsilon_{1,k} \le \varepsilon_{2, k} \le \cdots$ going to
infinity. The functions $k \mapsto \varepsilon_{n,k}$ are continuous
and $\cR'$-periodic. We assume in the sequel that there is a gap
$g > 0$ such that
\[
    { \forall k \in \cB, \quad \varepsilon_{N+1, k} - \varepsilon_{N,k} \geq g}
\]
where $N \in \N^*$ is the number of electrons per unit cell. In this
case, the operators $H(k)$ have a spectral gap, and we can define the
projector $P(k) := \1(H(k) \le \varepsilon_{N,k} +g/2)$. This
projector is of rank $N$, it is smooth with respect to $k$ and satisfies
the quasi-periodic conditions
\begin{align*}
  P(k+K) = \tau_K P(k) \tau_K^*.
\end{align*}
The projector on the occupied states $P$ is the operator acting on
$L^2(\C^d, \C)$, whose Bloch fibers are $P(k)$.

\subsection{Bloch frames and localisation of Wannier functions}

We say that $\bu(k) := (u_1(k), \cdots , u_N(k)) \in \left(L^2(\cA, \C) \right)^N$ is a Bloch
frame for $P(k)$ if, for all $k \in \R^d$, $\bu(k)$ is an orthonormal family spanning the
range of $P(k)$, and if
$\bu(k + K) = \tau_K \bu(k)$ for all $K \in \cR^*$. The Wannier
functions are then defined for $1 \le n \le N$ and $R \in \cR$ as
\begin{equation} \label{eq:wn_phin}
    w_{n,R}(x) := \dfrac{1}{| \cB |} \int_\cB \re^{ \ri k \cdot (x-R)} u_n(k,x) \rd k.
\end{equation}
We have $w_{n,R}(x) = w_{n,0}(x-R)$. Moreover, as the family
$\{u_n(k)\}_{1 \le n \le N}$ is an orthonormal basis of $\Ran P(k)$,
the family $\{ w_{n,R}\}_{1 \le n \le N, R \in \cR}$ is orthonormal in
$L^2(\R^d, \C)$, and spans the range of $P$. Finally, if furthermore
the map $k \mapsto u_n(k)$ is smooth, then the functions $w_{n,R}$ are
localised, as can be seen by integrating by part~\eqref{eq:wn_phin}.

We deduce that the existence of localised Wannier functions is
equivalent to the existence of a smooth frame $\bu$ for $P$. In other
words, we have reduced the problem of constructing localised Wannier
functions to that of the following problem: {\em given a smooth map of rank-$N$ projectors $k \in \R^d \mapsto P(k)$
satisfying $P(k+K) = \tau_K P(k) \tau_K^*$ for $K \in \cR^*$, can we
find a smooth frame $\bu(k)$ for $P(k)$ which satisfies
$\bu(k+K) = \tau_K \bu(k)$?}

 \subsection{Symmetries and topology}
 
 The existence of smooth Bloch frames (and therefore, of localised
 Wannier functions) in dimension $d \geq 2$ is not automatic, and
 depends on the topological properties of the Bloch bundle
 \cite{Brouder2007exp}. In dimension 2 and 3, the existence of
 localised Wannier functions is equivalent to the vanishing of
 topological invariants known as Chern numbers (one number in
 dimension $d=2$, and three numbers in dimension $d=3$).
 
In the important case where the map $k \mapsto P(k)$ satisfies the extra time-reversal symmetry (TRS), that is
\begin{equation} \label{eq:def:TRS}
    P(-k) = \theta P(k) \theta^{-1}, \quad \text{with} \quad \theta \ {\rm antiunitary},
\end{equation}
then these Chern numbers always vanish, and a smooth frame, together
with its corresponding localised Wannier functions, always
exists~\cite{nenciu1983existence, panati2007triviality}.

In the context of Schrödinger operators, condition~\eqref{eq:def:TRS} is satisfied
with $\theta u := \overline{u}$ the complex conjugation operator. This
operator is of bosonic type, squaring to $1$. If we start instead of
$H = -\frac 1 2 \Delta + V$ with a Hamiltonian including spin-orbit
coupling, we obtain a TRS of fermionic type, with an operator $\theta$
squaring to $-1$. In the case of FTRS, it is not always possible to
build localised Wannier functions that additionally respect a natural
symmetry condition \cite{Fiorenza2016}, causing many natural
algorithms to fail.


\begin{remark} \label{rem:smoothness}
    The existence of a smooth and quasi-periodic Bloch frame is a
    topological property. A consequence of the topological nature of the
    problem for our purposes is that, provided sufficient regularity on
    $k \mapsto H(k)$, if a continuous and quasi-periodic Bloch frame
    exists, then it can be lifted to a smooth and quasi-periodic one.
    Hence, in what follows, we will restrict ourselves to constructing
    continuous frames, as this can be regularised later, theoretically by
    the arguments in \cite{fiorenza2016construction} and numerically by
    using the Marzari-Vanderbilt procedure  \cite{marzari2012maximally}.
\end{remark}

\section{From Bloch frames to homotopies}
\label{sec:bloch_to_homotopies}

\subsection{Parallel transport}

An important notion that we use throughout the proof is {\em parallel transport}. We recall in this section the main properties of parallel transport, and explain how to solve it numerically.

Consider a smooth family of orthogonal projectors
$[0,1] \ni t \mapsto {P}(t)$, where $P(t)$ is a rank-$N$ projector
acting on some Hilbert space $\cH$. Let
$\bu(0) = (u_1(0), \ldots, u_m(0)) \in \cH^m$ be any set of $m$ vectors in $\Ran P(0)$, with $m \leq N$. 
Then the solution to the ordinary
differential equation
\begin{equation} \label{eq:parallel_transport}
    \bu'(t) = \left[ P'(t), P(t) \right] \bu(t), \quad \text{with} \quad  \bu(t = 0) = \bu(0)
\end{equation}
satisfies
\[
    (\bu^* \bu)'(t) = (\bu^*)'(t) \bu(t) + \bu^*(t) \bu'(t)
         = \bu^* \left( - \left[ P'(t), P(t) \right] + \left[ P'(t), P(t) \right] \right) \bu = 0
\]
and
\begin{align*}
    \left( \bu^* P \bu \right)'(t) & = \bu^*(t) \left( - \left[ P'(t), P(t) \right] P(t) + P'(t) + P(t) \left[ P'(t), P(t) \right]\right) \bu(t) \\
    & = \bu^*(t) \left( -P'(t) P + P(t) P'(t) P(t) + P'(t) + P(t) P'(t) P(t) - P(t) P'(t) \right) \bu(t) \\
    & = 0,
\end{align*}
where we used the fact that $P^2(t) = P(t)$, which first leads to
$P(t) P'(t) + P'(t) P(t) = P'(t)$, then to $P(t) P'(t) P(t) = 0$. It
follows that $u(t)$ is an orthonormal set of vectors in $\Ran P(t)$ for all $t \in [0,1]$. In particular, one can simplify~\eqref{eq:parallel_transport} with
 \[
     u'(t) = P'(t) P(t) u(t) - P(t) P'(t) u(t) = P'(t) u(t) - P(t) P'(t) P(t) u(t) = P'(t) u(t),
 \]
where we used the fact that $P(t) u(t) = u(t)$, and again the equality $P(t) P'(t) P(t) = 0$.
This gives the following orthogonality-preserving discretisation scheme. We assume that we are
given $P(t_i)$ for $0 = t_0 \le t_1 \le \cdots \le t_I = 1$, and
$u(0)$ an orthonormal family in the range of $P(0)$. Then we set
\begin{equation} \label{eq:numerical_parallel_transport}
     \begin{cases} 
       \widetilde{\bu}_{t_{i+1}} & = P(t_{i+1}) \bu_{t_{i}}, \\
       \bu_{t_{i+1}} & = \widetilde{\bu}_{t_{i+1}} \left[ \widetilde{\bu}_{t_{i+1}}^* \widetilde{\bu}_{t_{i+1}}\right]^{-1/2}.
   \end{cases}
 \end{equation}
 This is a convergent discretisation of \eqref{eq:parallel_transport},
 in the sense that when the spacing $\sup_{i} t_{i+1} - t_{i}$
 converges to zero, $\bu_{t_{i}}$ converges to $\bu(t_{i})$. 


\subsection{Obstruction matrices and homotopy}
In this section, we explain how to reduce the problem of constructing
a smooth Bloch frame in $d$ dimensions to that of finding a
$(d-1)$-homotopy of unitary matrices in $U(N)$. This is a standard
construction that was used in several articles (for instance,
\cite{cances2017robust,soluyanov2012smooth} and references therein). We
proceed by induction on the dimension $d = 1, 2, 3$.

%
%
%

\subsubsection{Construction for $d=1$}

In dimension $d = 1$, we are given a smooth family of projectors
$P(k_1)$ with $k_1 \in [0, 1]$, which satisfies the quasi-periodic
condition $P(1) = \tau_1 P(0) \tau_1^*$. We choose an arbitrary
orthonormal basis $\widetilde u(0)$ of $\Ran P(0)$. We then use
parallel transport~\eqref{eq:parallel_transport} to construct a smooth
frame $\widetilde{\bu}(k_1)$ for $P(k_1)$, for all $k_1 \in [0,1]$.
The problem is that $\widetilde{\bu}(1)$ is not equal to
$\tau_1 \widetilde{\bu}(0)$ {\it a priori}. Still, they both form an
orthonormal basis of $\Ran P(1) = \Ran P(0)$, and therefore are
related by a unitary matrix $V_\obs \in \U(N)$, called the
obstruction matrix:
\[
  \widetilde{\bu}(1)=  (\tau_1 \widetilde{\bu}(0))  V_\obs
\]
Since $V_\obs \in U(N)$, there is a anti-hermitian matrix $L$ such
that $V_\obs = \exp(L)$. We then set
\[
  \label{eq:1D_rectification}
    \bu(k_1) := \widetilde{\bu}(k_1) \exp(-k_1 L).
\]
By construction, $k_1 \mapsto \bu(k_1)$ is smooth on $[0, 1]$, and
satisfies $\bu(1) = \tau_1 \bu(0)$ as wanted. The continuous map
$k_{1} \mapsto \bu(k_{1})$ can be smoothed out following Remark~\ref{rem:smoothness}.
This gives the desired Bloch frame for $d = 1$.

%

\subsubsection{Construction for $d=2$}

The construction in two dimensions relies on the previous one-dimensional construction. We assume that we are given a smooth family $[0, 1]^2 \ni (k_1, k_2)  \to P(k_1, k_2)$ of operators satisfying $P(k + K) = \tau_K P(k) \tau_K^*$ for all $K \in \cR^*$.

First, we use the previous $d = 1$ construction
on the segment $k_2 = 0$, and get a smooth and quasi-periodic frame $\widetilde \bu(k_1, 0)$ for the family of projectors
$[0,1] \ni k_1 \to P(k_1, 0)$. Now for every
$k_{1} \in[0,1]$, we parallel transport the frame $\bu(k_1,0)$ along
the second direction, to produce a frame $\widetilde{\bu}(k_1, k_2)$
on $[0, 1]^2$. The frame is continuous, and satisfies
$\widetilde{\bu}(1, k_2) = \tau_{(1,0)} \widetilde{\bu}(0, k_2)$ for
all $k_2 \in [0,1]$. However, there may be a mismatch on the
$k_2$-boundary: for all $k_1 \in [0, 1]$, there is $V_\obs(k_1)$ so
that
\[
   \widetilde{\bu}(k_1, 1) = (\tau_{(0,1)} \widetilde{\bu}(k_1, 0)) V_\obs(k_1)
\]
In addition, since
$\widetilde{\bu}(1, 0) = \tau_{(1,0)} \widetilde{\bu}(0, 0)$ and
$= \widetilde{\bu}(1, 1) = \tau_{(1,0)} \widetilde{\bu}(0, 1)$, we have
$V_{\obs}(0) = V_\obs(1)$. The map $k \mapsto V_{\obs}(k)$ is
periodic, continuous and piecewise smooth on $\R$, and can be seen as a loop
$\mathbb{T}^{1} \to \U(N)$. We recall the following well-known fact.
\begin{proposition} \label{prop:homotopyd=2} Let
  $\mathbb{T}^{1} \ni k \mapsto V(k) \in \U(N)$ be a continuous and
  piecewise smooth loop in $\U(N)$. The two following assertions are
  equivalent:
     \begin{enumerate}
         \item The winding number $W(\det V)$ of the determinant of $V$ vanishes, where 
         \begin{equation} \label{eq:def:W}
            W(\det V) := \dfrac{1}{2 \pi} \int_0^1 \dfrac1{\det ( V(k) )} \det(V(k)) ' \rd k = \dfrac{1}{2 \pi} \int_0^1 \Tr(V^*(k) V'(k)) \rd k.
         \end{equation}
         \item There is a homotopy from $V(\cdot)$ to $\bbI_N$, that is a piecewise smooth map $ \TT^1 \times [0,1] \ni(k,t) \mapsto \homotopy{V}(k, t) \in \U(N)$ which satisfies
         \[
            \forall k \in \TT^1, \quad \homotopy{V}(k,0) = V(k) \quad \text{and} \quad \homotopy{V}(k,1) = \bbI_N.
         \]
     \end{enumerate}
\end{proposition}

In the next section, we give a constructive proof of this fact, in the sense that if the winding number vanishes, we construct algorithmically the homotopy $\homotopy{V}$. In our case, it can be further shown (see~\cite{fiorenza2016construction}) that $W(\det V_\obs)$ equals the Chern number of $P(k_1, k_2)$. According to this proposition, and assuming that $W(\det V_\obs) = 0$, there is a homotopy $\homotopy{V_{\obs}}(k_1, t)$ from $V_\obs$ to $\bbI_N$. We finally set
\[
    \bu(k_1, k_2) := \widetilde{\bu}(k_1, k_2) \homotopy{V_{\rm obs}}(k_1, k_2).
\]
By construction, this Bloch frame is continuous and satisfies the quasi-periodic boundary condition. It can be smoothed out following Remark~\ref{rem:smoothness}.

\subsubsection{Construction for $d=3$}
The extension to the third dimension is identical. 
First, use the $d=2$ procedure on the face $k_3 = 0$, \ie on $\{(k_1,k_2,0)\;,\; (k_1,k_2) \in [0,1]^2\}$, to obtain a Bloch frame $\widetilde{\bu}(k_1, k_2, 0)$ on this face. According to the previous section, this is possible if and only if the winding number of the obstruction on this face vanishes. Then, we parallel transport this frame along the third dimension and get $\widetilde{\bu}(k_1, k_2, k_3)$. We obtain another obstruction matrix $V_\obs(k_1, k_2) \in \U(N)$, such that
\[
    \forall k_1, k_2 \in [0, 1]^2, \quad \widetilde{\bu}(k_1,k_2,1) =   (\tau_{(0,0,1)}
    \widetilde{\bu}(k_1, k_2, 0)) V_\obs (k_1,k_2). 
\]
As before, we have $V_\obs(0, k_2) = V_\obs(1, k_2)$ and $V_\obs(k_1,
0) = V_\obs(k_1, 1)$, and so $V_\obs$ can be seen as a map from $\TT^2$ to $\U(N)$.
In the sequel, we prove the following classical result, which is the $2$-dimensional counterpart of Proposition~\ref{prop:homotopyd=2}
\begin{proposition} \label{prop:homotopyd=3}
    Let $\mathbb{T}^{2} \ni (k_1, k_2) \mapsto V(k_1, k_2) \in \U(N)$
    be a continuous and piecewise smooth surface in $\U(N)$. The two following assertions are equivalent:
    \begin{enumerate}
        \item The winding numbers $W( \det V(\cdot, 0))$ and $W(\det V(0, \cdot))$ both vanish;
        \item There is a $2$-homotopy from $V$ to $\bbI_N$, that is a smooth map $ \TT^2 \times [0,1] \ni(k_1, k_2, t) \mapsto \homotopy{V}(k_1, k_2, t) \in \U(N)$ which satisfies
        \[
        \forall k_1, k_2 \in \TT^2, \quad \homotopy{V}(k_1, k_2,0) =  V(k_1, k_2) \quad \text{and} \quad \homotopy{V}(k_1, k_2,1) = \bbI_N.
        \]
    \end{enumerate}
\end{proposition}
If the assertions are satisfied for our map $V_\obs(k_1, k_2)$, there is a $2$-homotopy $\homotopy{V_{\obs}}(k_1, k_2, t)$ that contracts $V_\obs$ to $\bbI_N$, and we set
\[
    \bu(k_1, k_2, k_3) := \widetilde{\bu}(k_1, k_2, k_3) \homotopy{V}_{\obs}(k_1, k_2, k_3)
\]
to obtain the final Bloch frame.

As in the $d=2$ case, the three winding numbers appearing in the
construction correspond to the three Chern numbers.


\begin{remark}
  This construction process extends trivially to dimensions $d > 3$,
  but the analogue of Propositions \ref{prop:homotopyd=2} and
  \ref{prop:homotopyd=3} are no longer true, and an additional
  obstruction (the second Chern class) appears.
\end{remark}

It remains to explain our constructive proof of Propositions~\ref{prop:homotopyd=2} and~\ref{prop:homotopyd=3}. This is the topic of the next section.

\section{Constructive homotopies in the unitary group}
\label{sec:homotopies_and_logs}

In this section, we describe a simple and efficient algorithm to
construct $1$-homotopies and $2$-homotopies in $\U(N)$. We first
examine how the logarithm algorithm in \cite{cances2017robust} fails
for simple systems such as the Kane-Mele model. We then explain our
algorithm in the context of $1$-homotopies, and then extend our
result for $2$-homotopies.

\subsection{Logarithm algorithm}

Let $\TT^1 \ni k \mapsto V(k) \in \U(N)$ be a smooth loop. A very
natural approach, that was used in~\cite{cances2017robust}, is to find
a global logarithm for $V(k)$, that is a smooth loop $L(k)$ of
anti-hermitian matrices such that
\[
    V(k) = \exp\left({L}(k) \right), \quad \forall  k\in [0,1].
\]
If such a logarithm exists, then a homotopy from $V(k)$ to $\bbI_N$ is given by
\[\forall k \in \TT^1, \forall t \in [0,1], \quad  \homotopy{V}(k,t) = \exp\left( (1-t) {L}(k) \right). \]
The authors of~\cite{cances2017robust} then proposed to work with the
eigenvalues of ${U}(k)$, to find a continuous phase for each on.
However, even if the winding number $W(\det V)$ vanishes, this
approach can fail, as shown by this simple example
\begin{example}
\label{exampleLog1}
Consider the analytic and periodic matrix path
\[V(k) =\begin{pmatrix}
\exp (2\ri\pi k) & 0 \\
0 & \exp(-2\ri\pi k)
\end{pmatrix} \]
Here, it is impossible to find a logarithm of the path that is continuous and periodic on $[0,1]$, since each eigenvalue has a winding number, hence receives a phase increment of $\pm 2\pi$ respectively when going from $0$ to $1$. 
\end{example}

\begin{figure}[!ht]
  \centering
  \begin{subfigure}{.5\textwidth}
  \centering
  \includegraphics[width=\linewidth]{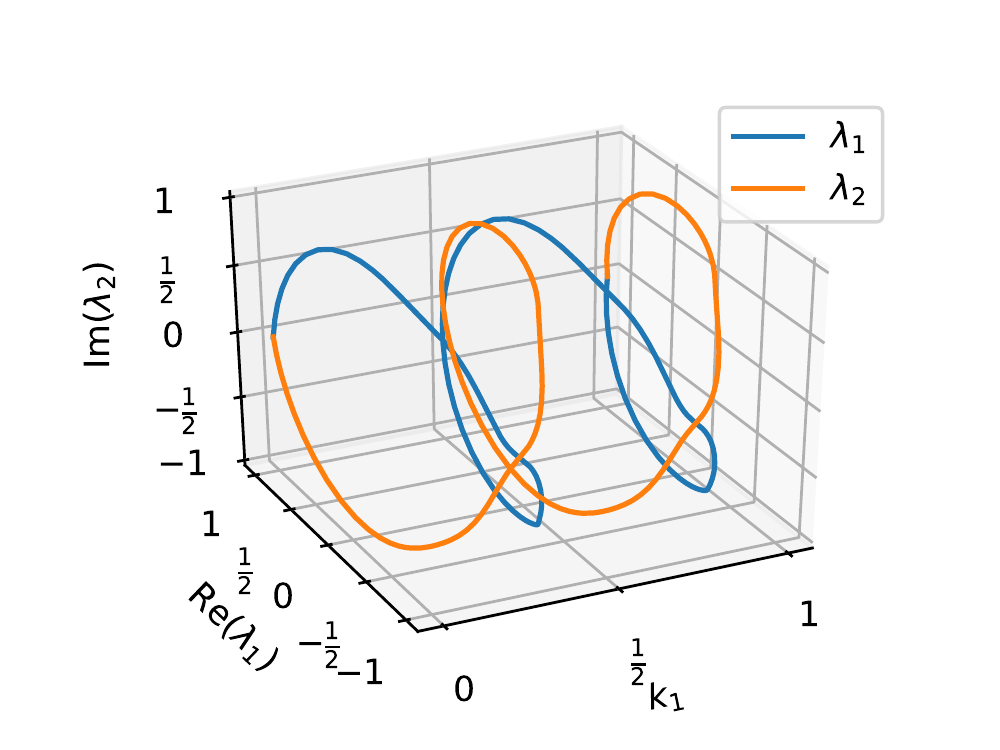}
  \caption{Eigenvalues winding in opposite directions}
  \label{eigObs}
  \end{subfigure}%
  \begin{subfigure}{.5\textwidth}
   \centering
  \includegraphics[width=0.893\linewidth]{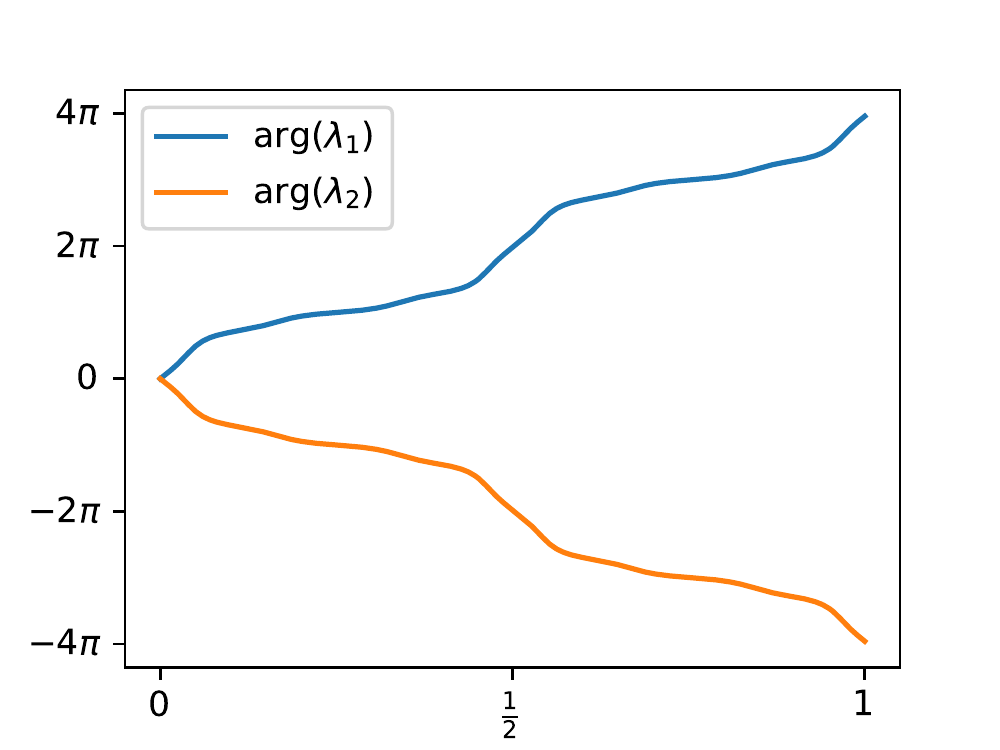}
  \caption{Phase of the eigenvalues, corrected for continuity}
  \label{logEigObs}
 \end{subfigure}
 \caption{Eigenvalues of the obstruction for the Kane-Mele model}
 \label{KaneMeleObs}
 \end{figure}

The case of eigenvalues having a winding number appears in practice
for systems with fermionic time-reversal symmetry such as the
Kane-Mele model in its QSH phase (see Section \ref{KaneMele}). In
Figure \ref{KaneMeleObs}, we display the eigenvalues of the
obstruction matrix for a representative set of parameters. Here, the
determinant is identically $1$. Hence, we know that a homotopy does
exist, but the logarithm method fails to construct it.

A similar method, proposed in \cite{cornean2017construction}, is to introduce a small perturbation in order to avoid eigenvalue crossings, which makes each winding number trivial, and look for a family of logarithms satisfying
\[{V}(k) = \mathrm{e}^{{L}_1(k)} \mathrm{e}^{{L}_2(k)} \dots \mathrm{e}^{{L}_N(k)},
\]
where ${L}_i(s), \; i \in 1\dots N$ are anti-Hermitian. 
However, small perturbations of eigenvalues can introduce large changes in the eigenvectors, 
and hence produce a continuous but irregular path, which makes
this method algorithmically difficult to implement.


\subsection{Column interpolation method}
\label{sec:ourMethod}

From the counter-example given in Example \ref{exampleLog1}, we see
that constructing a homotopy of unitary matrices based from their
eigenvalues may fail, as these can wind. In our method, instead of
contracting eigenvalues, we rather contract the columns of $V(k)$ one
by one. Algebraically, this corresponds to exploiting the fibration
\[\U(N-1) \to \U(N) \to \SS^{2N-1},\]
which was suggested (but not explored further)
in~\cite[p.81]{fiorenza2016construction}.

let $\TT^1 \ni k \mapsto V(k) \in \U(N)$ be a smooth map. We write
$V(k) = (v_1(k), \ldots , v_N(k))$ where $v_n(k) \in \SS^{2N-1}$ is
the $n$-th column of $V(k)$. Our strategy is to first contract the
columns $v_{n}(k)$ iteratively to a fixed column $\underline{v_{n}}$,
ensuring that they stay orthonormal, and then homotopise
$\underline{V} = (\underline{v_1}, \ldots , \underline{v_N})$ to
the identity.

Let us assume that at step $1 \le n \le N$, we have found how to
contract the first $n-1$ columns to some fixed vectors: we have
constructed $n-1$ smooth maps of orthonormal vectors $v_1(k, t), \ldots, v_{n-1}(k,t)$ such that $v_j(k,t=0)= v_j(k)$ and $v_j(k,t=1) = \underline{v_j}$. We denote by 
\[
    P_{n-1}(k,t) := \mathbb I_{N} -  \sum_{j=1}^{n-1} | v_j(k,t) \rangle \langle v_j(k,t) |,
\]
the projection on the orthogonal of this family, of rank $N-n+1$. By hypothesis, at $t = 1$, the projectors $P_{n-1}(k,t=1)$ are equal to a constant projector
$\underline{P_{n-1}}$.

We now contract the $n$-th column $v_{n}(k,t)$ to a fixed column
$\underline {v_{n}} \in \Ran \underline{P_{n-1}}$ while satisfying
$v_{n}(k,t) \in \Ran P_{n-1}(k,t)$ for all $k,t \in \TT^1 \cup [0,1]$. This ensures that the constructed map for the $n$-th column is orthogonal to the previously constructed ones.
First, for all fixed $k \in \TT^1$, we parallel transport the
orthogonal family $(v_n(k), \cdots, v_N(k))$ with respect to
$P_{n-1}(k, \cdot)$, and obtain a smooth family of orthonormal frames
$(\widetilde{v_n}(k,t), \cdots, \widetilde{v_N}(k,t))$ for
$k,t \in \TT^1 \times [0,1]$. At this point, $\widetilde{v_n}(k,t=1)$
forms a non-trivial loop in $\Ran \underline{P_{n-1}}$. We now contract this
to a single vector $\underline{v_n}$, distinguishing two cases,
depending on whether $\widetilde{v_n}(k,t=1)$ can or cannot cover the
whole of the unit sphere in $\Ran \underline{P_{n-1}}$.

\paragraph{Case $n < N$.}
When $n < N$, the unit sphere in $\Ran \underline{P_{n-1}}$ is a
manifold of real dimension $2(N-n)+1 \ge 3$. The family
$\left\{\widetilde{v_n}(k,t=1) \right\}_{k \in \TT^1}$ describes a
piecewise smooth loop on that manifold, and from Sard's lemma it
follows that there exists a vector
$\underline{v_n} \in \SS^{2N-1} \cap \Ran \underline{P_{n-1}}$ such
that $-\underline{v_n}$ does not belong to the loop
$\left\{\widetilde{v_n}(k,t=1) \right\}_{k \in \TT^1}$ (see Remark
\ref{rem:find_fixed_vector}).

For all $k \in \TT^1$, the family
$(\widetilde{v_n}(k,1), \cdots, \widetilde{v_N}(k,1))$ is a basis of
$\underline{P_{n-1}}$, so there exist (smooth) coefficients
$c(k) := (c_n(k), \ldots, c_{N}(k)) \in \C^{N-n+1}$ with $|c(k)|=1$ such that
\[
    \forall k \in \TT^1, \quad \underline{v_n} = \sum_{j=n}^N c_j(k) \widetilde{v_j}(k,1).
\]
The map $\TT^1 \ni k \mapsto c(k)$ is a loop on the sphere
$\SS^{2(N-n)+1}$. In addition, since $-\underline{v_n}$ never touches
the loop $\left\{\widetilde{v_n}(k,t=1) \right\}_{k \in \TT^1}$,
$c(k)$ never touches the vector $-e_1 := (-1, 0,\cdots, 0)$. We can
therefore contract the loop $c(k)$ to $e_1$ on $\SS^{2(N-n)+1}$ with
the explicit contraction
\begin{equation} \label{eq:def:contraction_for_c}
\homotopy{c}(k,t) := \dfrac{(1-t) c(k) + t e_1}{\| (1-s) c(k) + t e_1\|},
\end{equation}
which is a well-defined smooth map from $\TT^1 \times [0,1]$ to $\SS^{2(N-n)+1}$. This contraction of coefficients directly translates into a contraction of $v_n(k)$ to $\underline{v_n}$ by setting
\[
    v_n(k,t) := \sum_{j=n}^N c_j(k,t) \widetilde{v_j}(k,t).
\]
By construction, $v_n(k,t)$ is a normalised vector which is orthogonal to $(v_1(k,t), \ldots, v_{n-1}(k,t))$ for all $k,t \in \TT^1 \times [0,1]$. This concludes the construction in this case.

\begin{remark} \label{rem:find_fixed_vector}
    In practice, in order to find numerically $\underline{v_n}$, we
    draw several random or well-chosen points $p_j \in \SS^{2N-1}$, which we project on $\underline{P_{n-1}}$ and normalise. We then pick 
    \[
    \underline{v_n} := \argmax_{j} \min_{k \in \TT^1} \| \widetilde{v_n}(k, 1) + p_j \|.
    \]
    This ensures that the denominator in~\eqref{eq:def:contraction_for_c} is not too close to zero.
\end{remark}

\paragraph{Case $n = N$.}
For the last vector, \ie when $n = N$, the previous construction can
fail because $\widetilde {v_{N}}(k,t=1)$ can cover the whole of the
unit sphere in $\Ran \underline{P_{n-1}}$. We therefore follow a
different route. For all $k \in \TT^1$, the vector $\widetilde{v_N}(k,t=1)$ always lies in the same one-dimensional subspace. In particular, there is a smooth phase $\phi : [0,1] \to \R$ so that 
\[
\forall k \in [0,1], \quad \widetilde{v_N}(k) = \underline{v_N} \re^{\ri \phi(k)} \quad \text{with} \quad \underline{v_N} := \widetilde{v_N}(0) \quad \text{(for instance)}.
\]
By periodicity, one must have $\phi(1) = \phi(0) + 2 \pi m$ with $m
\in \ZZ$. This gives
\begin{align*} 
m & = \frac{1}{2 \pi} \left( \phi(1) - \phi(0) \right) = \dfrac{1}{2 \pi} \int_0^1 \phi'(k) \rd k
= \dfrac{1}{2 \pi \ri} \int_0^1 \left\langle \widetilde{v_N}(k),  \frac{\rd}{\rd k}\widetilde{v_N}(k)  \right\rangle \rd k.
\end{align*}
We set $\widetilde{V}(k,t) := (v_1(k,t), \ldots v_{N-1}(k,t) ,
\widetilde{v_N}(k,t) ) \in \U(N)$. This is a smooth deformation between
$V(k)$ at $t=0$ and $\widetilde V(k,1)
= (\underline{v_1}, \ldots \underline{v_{N-1}} ,
\widetilde{v_N}(k,1) )$ at $t=1$. Also, we have
\[
    \left\langle \widetilde{v_N}(k),  \frac{\rd}{\rd k}\widetilde{v_N}(k)  \right\rangle = 
    \Tr \left( \widetilde{V}(k, 1)^* \frac{\rd}{\rd k} \widetilde{V}(k, 1) \right) \rd k.
\] 
This leads to
\begin{equation} \label{eq:def:m}
 m = \dfrac{1}{2 \pi \ri} \int_0^1 \Tr \left( \widetilde{V}(k, 1)^* \frac{\rd}{\rd k} \widetilde{V}(k, 1) \right) \rd k 
 = W\left( \det \widetilde{V}(\cdot,1)\right) = W( \det V(\cdot)),
\end{equation}
where we recall that $W(\cdot)$ was defined in~\eqref{eq:def:W}, and
where we used the fact that the winding number is not affected by a
smooth deformation: $W\left(\widetilde V(\cdot, t)\right)$ does not depend on
$t$. We conclude that can contract the vector $\widetilde{v_N}$ to
$\underline{v_N}$ if and only if $m = 0$, or equivalently if
$W(\det V) = 0$. In this case, an explicit contraction is given by
\[
    v_N(k,t) = \widetilde{v_N}(k,t) \re^{- \ri t \phi(k)}.
\]

\paragraph{Last step.} At this point, we have algorithmically constructed a smooth map $\TT^1 \times [0,1] \ni (k,t) \mapsto V(k,t) \in \U(N)$ such that $V(k,t=0) = V(k)$ and $V(k,t = 1) = \underline{V} := (\underline{v_1}, \cdots, \underline{v_N} )$. To get a contraction to the identity matrix $\bbI_N$, we write $\underline{V} = \exp (\underline{L})$, where $\underline{L}$ is anti-hermitian, and we take as our final homotopy 
\[
    (k,t) \mapsto V(k,t) \re^{-t \underline{L}}.
\]
This concludes our constructive proof for Proposition~\ref{prop:homotopyd=2}.

\begin{remark}
  In our algorithm, we have tried to make the homotopy as smooth as
  possible. This means that we avoid composing homotopies
  sequentially, which is inefficient numerically, and that we wish
  that the method reduces to the logarithm method in the case where
  $V(k)$ is constant (where we know that the logarithm gives the
  geodesic in $\U(N)$ and therefore the most efficient path). If that
  is not a concern, then a simpler version of the algorithm can be
  given. After the first column is homotopised to a column
  $\underline{v_{1}}$, this vector can further be deformed to $e_{1}$,
  and therefore we can assume that $\underline{v_{1}} = e_{1}$. This
  implies that the homotopy
  $\widetilde{V}(k,t) := (v_1(k,t), \widetilde{v_2}(k,t), \ldots,
  \widetilde{v_N}(k,t)))$ satisfies $\homotopy{V}(k,0) = V(k)$ and
\[
    \widetilde{V}(k, 1) = \begin{pmatrix}
    1 & \widetilde{v_{1,2}}(k,1) & \cdots &  \widetilde{v_{1,N}}(k,1)  \\
    0 &  \widetilde{v_{2,2}}(k,1) & \cdots &  \widetilde{v_{2,N}}(k,1) \\
    \vdots & \vdots & \cdots & \vdots \\
    0 &  \widetilde{v_{2,N}}(k,1) & \cdots & \widetilde{v_{N,N}}(k,1)
    \end{pmatrix}
    =:
     \begin{pmatrix}
    1 & 0   \\
    0 &  V_1(k)
    \end{pmatrix},
\]
where we used the fact that $\widetilde{V}(k, 1)$ is unitary, so that
$\widetilde{v_{1,2}}(k,1) = \cdots = \widetilde{v_{1,N}}(k,1) = 0$. We
have reduced the homotopy problem in $\U(N)$ to the homotopy problem
in $\U(N-1)$, and therefore solve the problem by induction on $N$,
using the case $n=N$ above to treat the base step. 
\end{remark}

\begin{remark}[Parallelisability of the sphere]
  In the case $N=2$, we can use the identification of $\SU(2)$ with
  $\SS(3)$ given by
  \begin{align*}
    \begin{pmatrix}
      a&-b^{*}\\
      b&a^{*}
    \end{pmatrix} \mapsto
    \begin{pmatrix}
      a\\b
    \end{pmatrix}
  \end{align*}
  to simplify the algorithm, as done in
  \cite{fiorenza2016construction}. More generally, if given a vector
  $x \in \{z \in \C^{N}, |z|=1\}$ we had a systematic way to build an
  orthogonal basis of the (complex-)orthogonal $x^{\perp}$ in a way
  that is smooth with respect to $x$, we could exploit that in our
  algorithm. This is easily achieved in dimension $2$ by the mapping
  $(a,b) \mapsto (-b^{*}, a^{*})$. However, this is impossible when
  $N=3$ (because this would imply the parallelisability of the
  $5$-dimensional sphere, which is false). We therefore have to follow
  a different route, using parallel transport to build this basis
  incrementally.
\end{remark}

\subsubsection{Extension for $2$-homotopies}
We now consider the case of $2$-homotopies, and we want to contract a
map $\TT^2 \ni (k_1, k_2) \mapsto V(k_1, k_2) \in \U(N)$. Following
the same iterations as in the previous section, we see that at step $n
< N$, the $n$-th column $\widetilde{v}(k_1, k_2, t=1)$ defines a
$2$-dimensional sub-manifold on $\SS^{2N-1} \cap \Ran \underline{P_{n-1}}$ of dimension $2(N-n)+1 \ge 3$, and we can find $\underline{v_n}$ so that $\underline{v_n}$ does not belong to this sub-manifold. We then follow the same steps.

For the last step $n = N$, there is a smooth phase function $\TT^2 \ni (k_1, k_2) \mapsto \phi(k_1, k_2)$ such that
\[
    \forall k_1, k_2 \in [0,1]^2, \quad \widetilde{v_N}(k_1, k_2, 1)= \underline{ v_N } \exp ( \ri \phi(k_1, k_2)) \quad \text{with} \quad 
    \underline{ v_N } := \widetilde{v_N}(0, 0, 1) \quad \text{for instance}.
\]
By periodicity and continuity, there is $m_1, m_2 \in \Z$ such that
$\phi(k_1 + 1, k_2) = \phi(k_1, k_2) + 2 \pi m_1$ and $\phi(k_1, k_2 +
1) = \phi(k_1, k_2) + 2 \pi m_2 $. As in~\eqref{eq:def:m}, we find
\[
    \forall k_2 \in \TT^1, \quad m_1 = W( \det V(\cdot , k_2)) \quad \text{and} \quad \forall k_2 \in \TT^1, \quad m_2 = W( \det V(k_1 , \cdot)).
\]
If both number vanish, then a contraction is given by $\homotopy{v_n}(k_1, k_2, t) := \widetilde{\homotopy{v_n}}(k_1, k_2, t) \exp(- \ri t \phi(k_1, k_2))$. The constructive proof of Proposition~\ref{prop:homotopyd=3} follows.

\begin{remark}
    This proof fails for $3$-homotopies. The reason is that with $N = 2$, the first vector of $\TT^3 \ni (k_1, k_2, t_3) \mapsto V(k_1, k_2, t_3) \in \U(2)$ is now a $3$-dimensional sub-manifold in $\SS^3$, hence can wrap the whole sphere $\SS^3$. This is a manifestation of the second Chern class.
\end{remark}

\section{Numerical results}
    \label{sec:num_res}
    In this section, we apply the constructive method outlined above
    to the case of the Kane-Mele model ($d=2$), and silicon ($d=3$).
    We discretise the Brillouin zone with equispaced points (the
    Monkhorst-Pack grid). At each discrete point $k$, we diagonalise
    $H(k)$ and obtain the eigenvectors $(u_{n,k})_{1\leq n \leq N}$
    corresponding to the $N$ lowest eigenvalues of $H(k)$. We then
    seek a unitary matrix $U_{m,n}(k)$ that makes
    $u'_{n}(k) = \sum_{1 \leq m \leq N} u_{mk}U_{mn}(k)$ as smooth as
    possible. The quantities needed by our algorithm are the overlaps
    $\langle u_{mk},u_{n,k+b} \rangle$ between neighbouring points $k$
    and $k+b$, similar to other methods such as Wannier90
    \cite{Wannier90}. More information on this methodology can be
    found in \cite{cances2017robust}.


\subsection{The Kane-Mele model}
\label{KaneMele}

The Kane-Mele model, first proposed in \cite{kane2005z}, is a toy
model of a $\mathbb{Z}^2$ topological insulator. It is a tight-binding
model on a 2D hexagonal lattice, with four degrees of freedom per site
(two orbitals and two spins), two of which are occupied ($H(k)$ is a
$4 \times 4$ matrix, and $N=2$).

\subsubsection{Description of the model}

The Bloch representation of this model can be written as follows.
\begin{equation} \label{eq:def:KaneMele}
    H_{k} = \sum_{a=1}^5 d_a({k}) \Gamma^a + \sum_{a,b = 1 \atop a < b}^5 d_{ab}({k})\Gamma^{ab},
\end{equation}
 where $\Gamma^{ab} := \frac{1}{2\ri}[\Gamma^a,\Gamma^b]$, and  $\Gamma^a$ are
 the Dirac matrices $(\sigma^x\otimes\bbI_N,\sigma^z \otimes\bbI_N,
 \sigma^y\otimes {s}^x,\sigma^y\otimes {s}^y, \sigma^y\otimes {s}^z)$,
$\sigma^j$ and  ${s}^j$ being the Pauli matrices of sublattice and spin respectively.


The functions $d_a(k)$ and $d_{ab}(k)$ in~\eqref{eq:def:KaneMele} are
chosen as in \cite{kane2005z}. In particular, $d_{a}$ is even and
$d_{ab}$ odd, and the model satisfies a fermionic time-reversal
symmetry. The model has $4$ free parameters: $t$, $\lambda_{S0}$,
$\lambda_\nu$ and $\lambda_R$. Here, we fix the parameters
$t=1,\lambda_{SO}=1$, and only vary $\lambda_\nu$ and
$\lambda_R<2\sqrt{3}$. For every value of $\lambda_{R} < 2\sqrt{3}$,
the system undergoes a phase transition at the critical value
$\lambda_{\nu} = 3 \sqrt{3} \approx 5.2$:
\begin{itemize}
\item For $\lambda_\nu > 3\sqrt{3}$, the material is in a regular insulating phase.
\item For $\lambda_\nu = 3\sqrt{3}$, the material is in a transitional metallic phase: the gap closes, which means that
the material is conducting.
\item For $\lambda_\nu < 3\sqrt{3}$, the material is in the Quantum Spin Hall (QSH) phase.
\end{itemize}

\subsubsection{Numerical construction of Wannier functions for the Kane-Mele model}

In order to construct localised Wannier functions for the Kane-Mele
model, one needs to provide a Bloch frame that is regular enough on
the Brillouin zone. In the QSH phase, no continuous and symmetric
frame exists, but since the Chern number is trivial for any
time-reversal symmetric Bloch bundle, there exists a non-symmetric
continuous frame. Moreover, in this case, the eigenvalues of the
obstruction have a non-trivial winding number, so the logarithm method
of \cite{cances2017robust} cannot provide a homotopy of the
obstruction.

\begin{figure}[!ht]
  \centering
   \begin{subfigure}{.5\textwidth}
   \centering
  \includegraphics[width=0.95\textwidth]{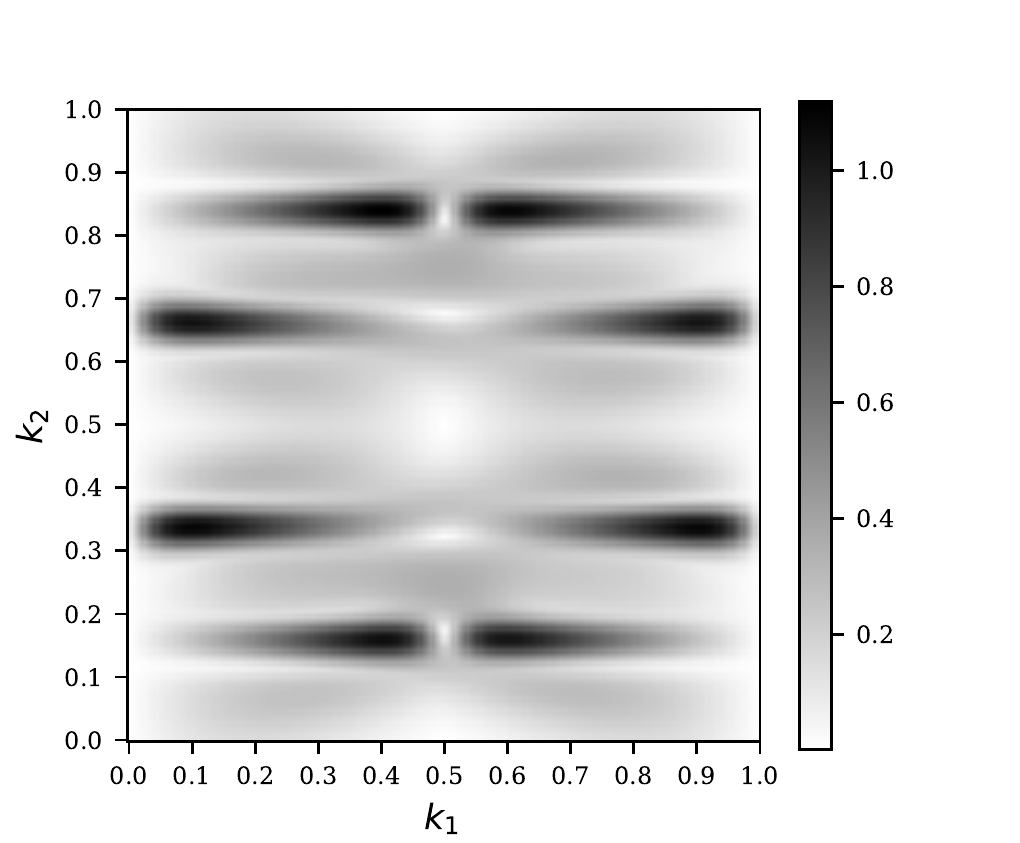}
  \caption{Logarithm interpolation}
  \label{insulatinglog}
\end{subfigure}%
 \begin{subfigure}{.5\textwidth}
  \centering
  \includegraphics[width=0.95\textwidth]{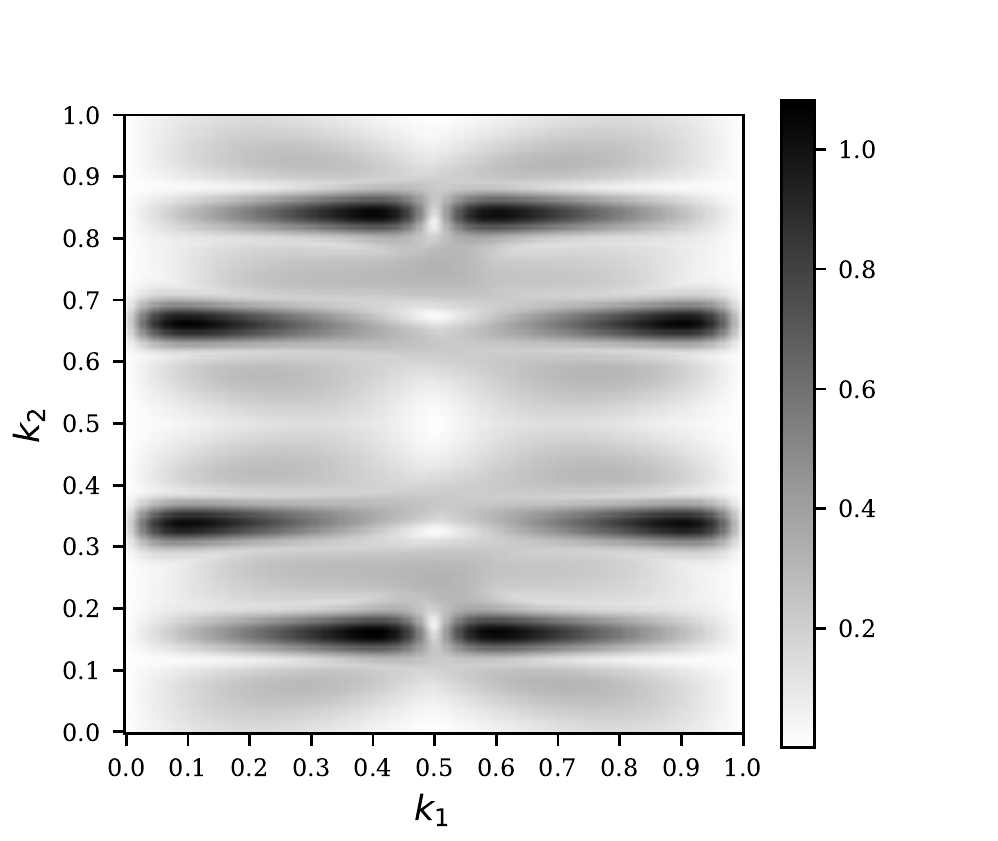}
  \caption{Column interpolation}
 \label{insulatingparallel}
 \end{subfigure}
 \centering
 \caption{Local regularity $\| \nabla_{k} {u}\|$,
   $\lambda_\nu = 6$ and $\lambda_R = 1$ (regular insulating phase).}
 \end{figure}

In this section, we use the algorithm described above to construct a
continuous initial guess of the Bloch frame, which can then be refined to a more regular one by a smoothing method, thus providing a well-localised Wannier function. The Brillouin zone was discretised with a $200\times200$ grid.
In the topologically trivial case, both methods produce a reasonable output (Figures \ref{insulatinglog} and \ref{insulatingparallel}).

In order to measure localisation, we follow~\cite{marzari1997maximally},
and measure the spread of the Wannier functions $\Omega$. We also
measure the quantity $\|\nabla_{k} u_{k}\|$, estimated using finite
differences. Localised Wannier functions correspond to smooth gauge,
and singularities in this quantity is therefore a sign of
delocalisation.

 \begin{figure}[!ht]
  \centering
   \begin{subfigure}{.5\textwidth}
   \centering
   \includegraphics[width=0.95\textwidth]{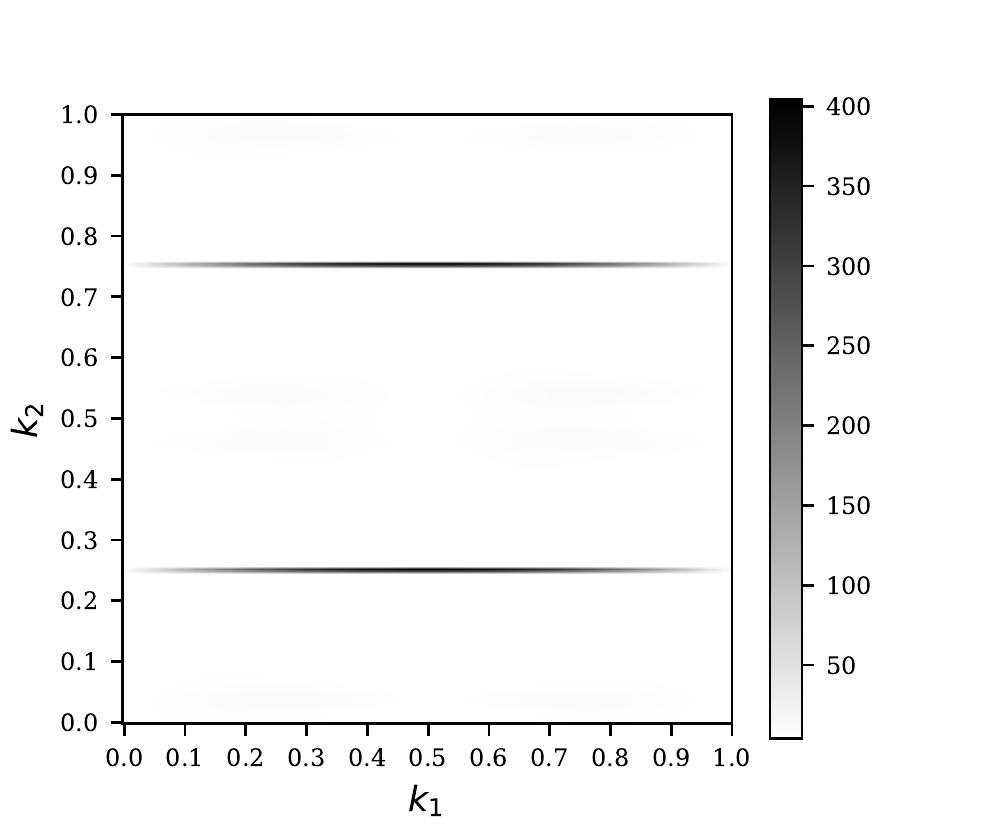}
  \caption{Logarithm interpolation}
  \label{QSHlog}
\end{subfigure}%
 \begin{subfigure}{.5\textwidth}
  \centering
  \includegraphics[width=0.95\textwidth]{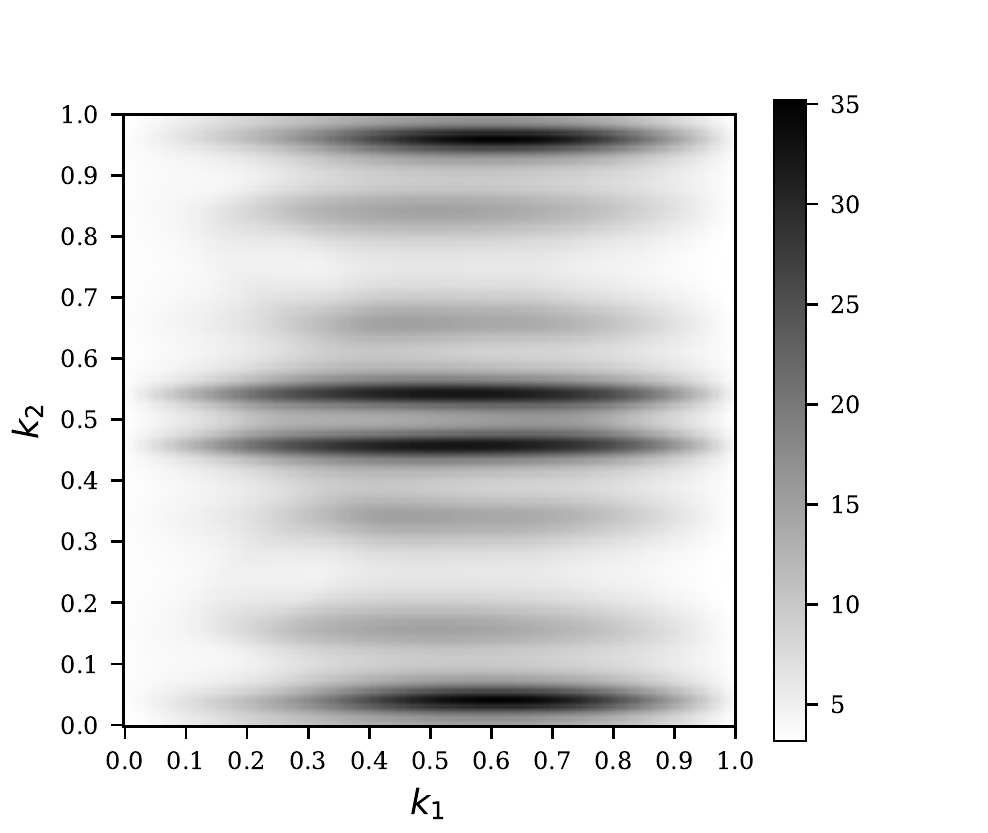}
  \caption{Column interpolation}
 \label{QSHparallel}
 \end{subfigure}
 \centering
 \caption{Local regularity $\| \nabla_{k} {u}\|$,
   $\lambda_\nu = 0$ and $\lambda_R = 1$ (QSH phase).}
 \end{figure}
 
In Figure \ref{QSHlog}, the log interpolation method fails at constructing a continuous map in the topologically non-trivial QSH phase, as the measure of regularity $\| \nabla_{k} {u}\|$ exhibits lines of discontinuity, with very high maximal values. 
In contrast, in Figure \ref{QSHparallel}, the column interpolation produces a smoother output, which also yields a lower maximal value of the regularity $\| \nabla_{k} {u}\|$.

 \begin{figure}[!ht]
  \centering
  \includegraphics[width=0.65\textwidth]{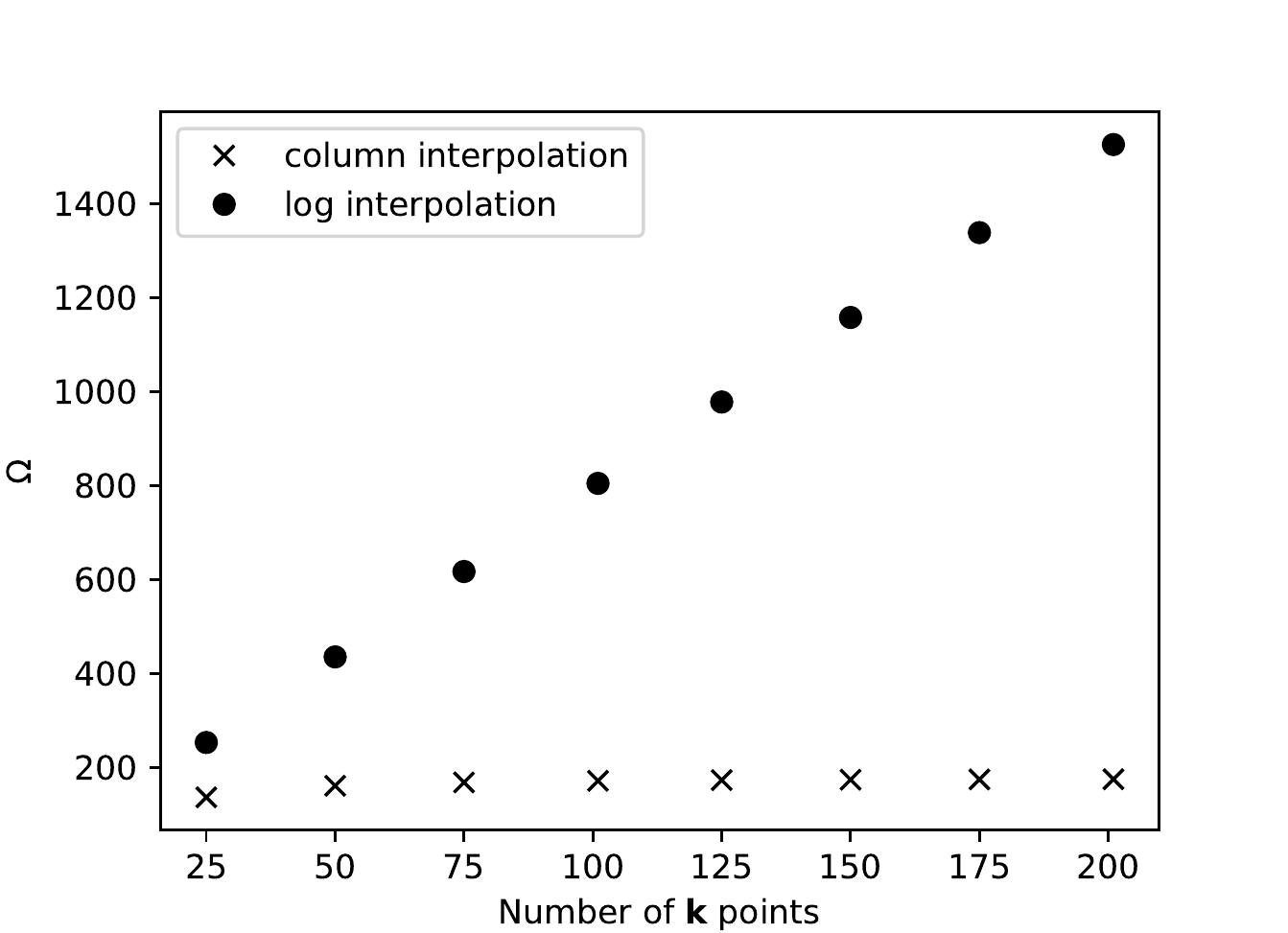}
  \caption{Convergence of $\Omega $ obtained by both methods, in the QSH phase ($\lambda_\nu = 0, \lambda_R = 1$)}
 \label{convergence_QSH}
 \end{figure}
The (dis)continuity of the resulting Bloch frame after each method is further demonstrated by the convergence with respect to ${k}$ point discretisation, displayed in Figure \ref{convergence_QSH}. In the log interpolation method, the discrete Bloch frames converge to a discontinuous one, as we see from the divergence of the norm of the gradient (estimated with finite differences). In contrast, the column interpolation produces a frame that has a smooth limit.

\begin{figure}[!ht]
\centering
 \begin{subfigure}{.5\textwidth}
   \centering
 \includegraphics[width=0.95\textwidth]{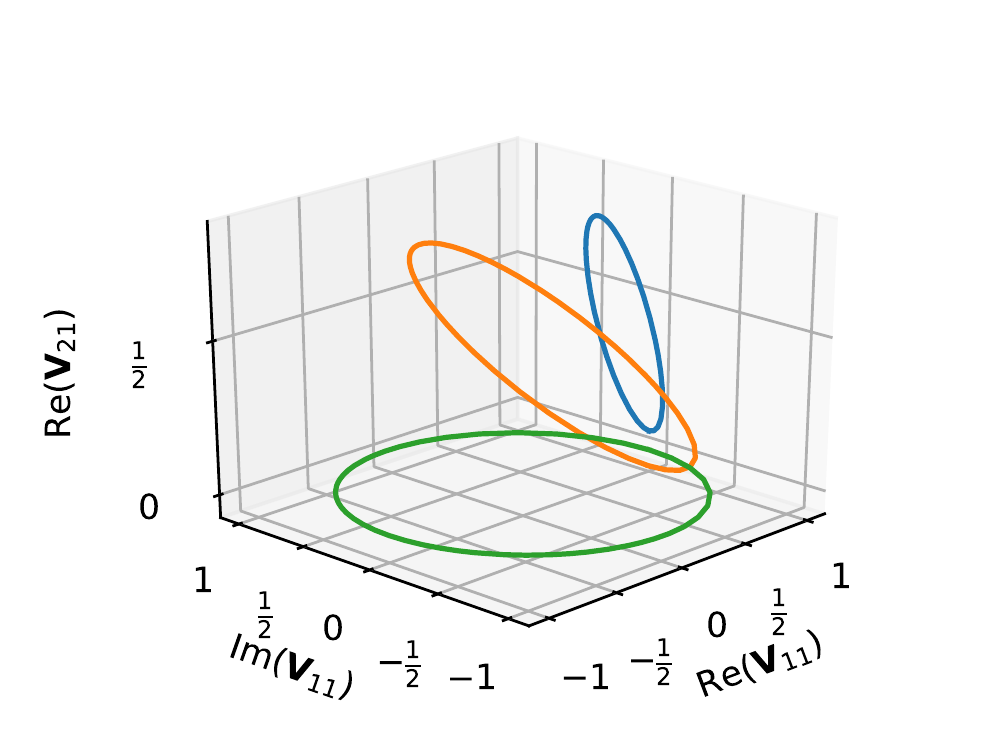}
  \caption{$\lambda_\nu = 1, \lambda_R = 0$}
  \label{QSHcontraction0}
 \end{subfigure}%
 \begin{subfigure}{.5\textwidth}
   \centering
    \includegraphics[width=0.95\textwidth]{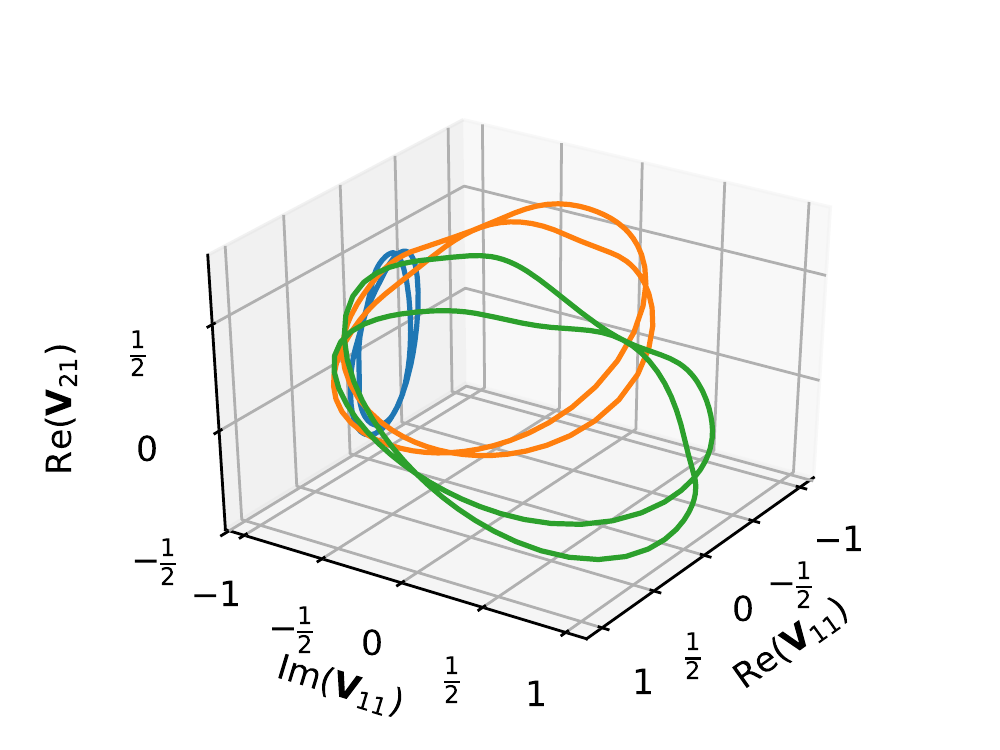}
  \caption{$\lambda_\nu = 1, \lambda_R = 1$}
  \label{QSHcontraction1}
  \end{subfigure}
  \caption{Contraction of the first column of the obstruction, with
    the initial path in green ($k_{2}=1$) being deformed to a single
    point (yellow at $k_{2}=\frac 2 3$ then blue at
    $k_{2} = \frac 1 3$).}
 \end{figure}

Figures \ref{QSHcontraction0} and  \ref{QSHcontraction1} display
selected components of $k_1\mapsto{V}_{\rm obs}(k_1,k_2)$ for $k_2=1,
\frac{2}{3}, \frac{1}{3}$. In Figure \ref{QSHcontraction0}, we see how the obstruction path is contracted into the null path by our algorithm, in the QSH phase, with no Rashba term. In this case, the system decomposes into two independent copies of the Haldane model, one for each spin, which implies that the obstruction matrix is diagonal. 
This explains that the obstruction (the largest path in the plot) is horizontal, since ${V}_{21}=0$.
Notice also that the diagonality of the obstruction, as well as time reversal symmetry, implying that ${k}\mapsto \Re V_{21}({k})$ is odd (which is verified up to rounding errors in the horizontal path), is broken by the method, as expected.

In Figure \ref{QSHcontraction1}, for a Rashba term $\lambda_R = 1$, the obstruction (the largest path, in green) is no longer diagonal (it has non-zero off-diagonal elements), but it still satisfies time-reversal symmetry, since ${k}\mapsto \Re V_{21}({k})$ is odd. The method breaks time-reversal symmetry to construct the continuous interpolation to the trivial path. 

\subsection{Numerical results for Silicon}

Using Quantum Espresso, \cite{QE-2009}, the Bloch waves of Silicon for
various discretisations of the Brillouin zone were provided to the
homotopy constructing methods, in order to compare the numerical
results of our column interpolation algorithm with the ones provided
by the logarithm method of \cite{cances2017robust}.

\begin{table}[!ht]
\caption{Value of the Marzari-Vanderbilt localisation functional $\Omega$ (in Bohr$^{2}$) for frames on various discretisations of the Brillouin zone}
\begin{center}
\begin{tabular}{|c|c|c|c|c|c|c|}
\hline
Discretisation of the BZ &   $5\times5\times5$ & $10\times10\times10$ & $15\times15\times15$ & $20\times20\times20$ \\
\hline
After logarithm method  &    25.72 &  29.70 & 30.62 & 30.94 \\
\hline
After column interpolation & 40.88 & 35.31 & 53.68 & 46.80 \\
\hline 
After MV optimisation  & & & & \\ 
(log initial guess) & 19.30 & 22.06 & 22.71 & 22.95 \\
\hline
After MV optimisation & & & & \\ 
(col initial guess) & 19.30 & 22.06 & 22.71 & 22.95 \\
\hline 
\end{tabular}
\end{center}
\label{locSilicon}
\end{table}

\begin{figure}[!ht]
  \centering
  \begin{subfigure}{.5\textwidth}
  \centering
  \includegraphics[width=0.8\linewidth]{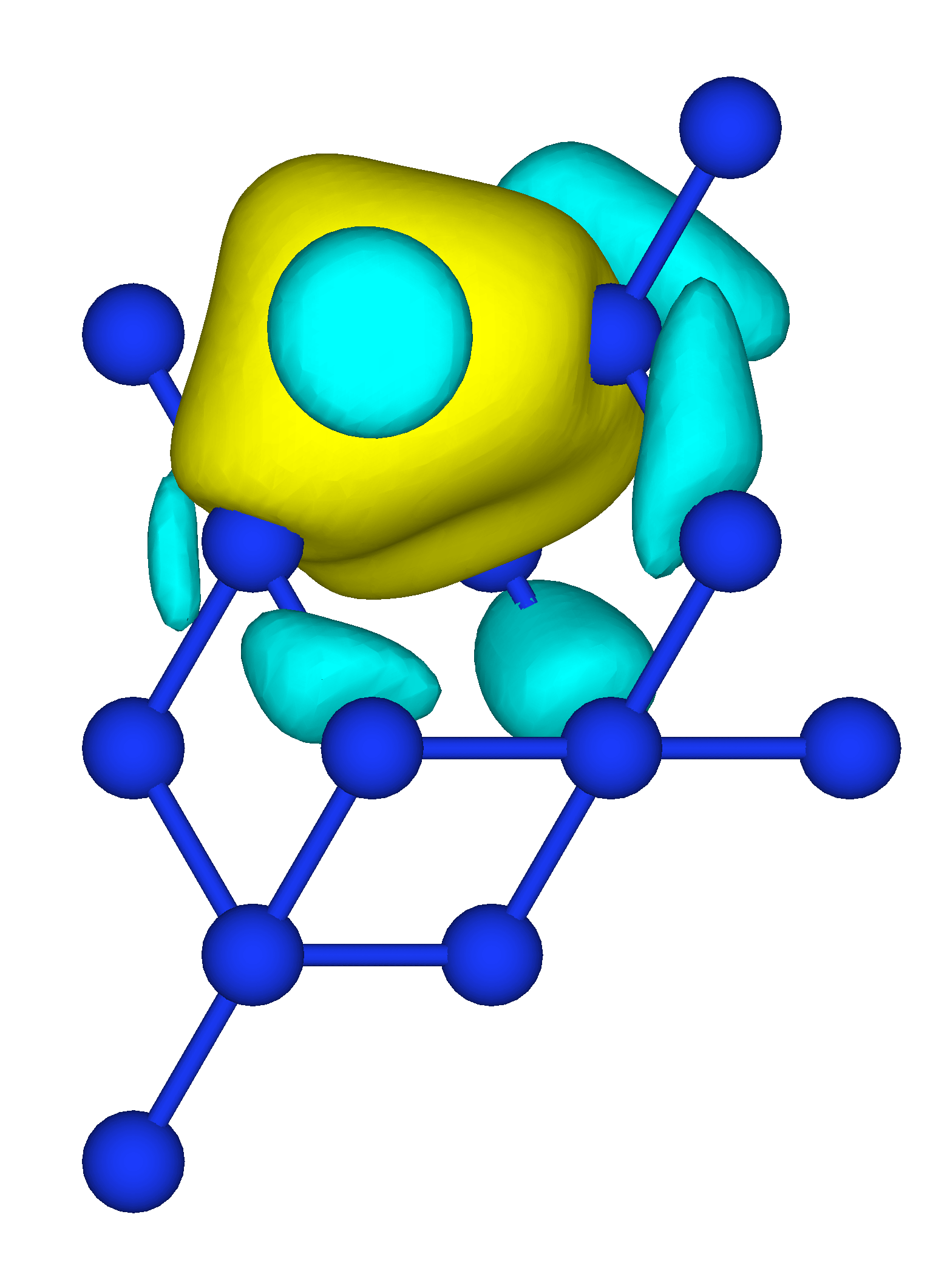}
  \caption{Logarithm method}
  \label{log_silicon}
  \end{subfigure}%
  \begin{subfigure}{.5\textwidth}
   \centering
  \includegraphics[width=0.8\linewidth]{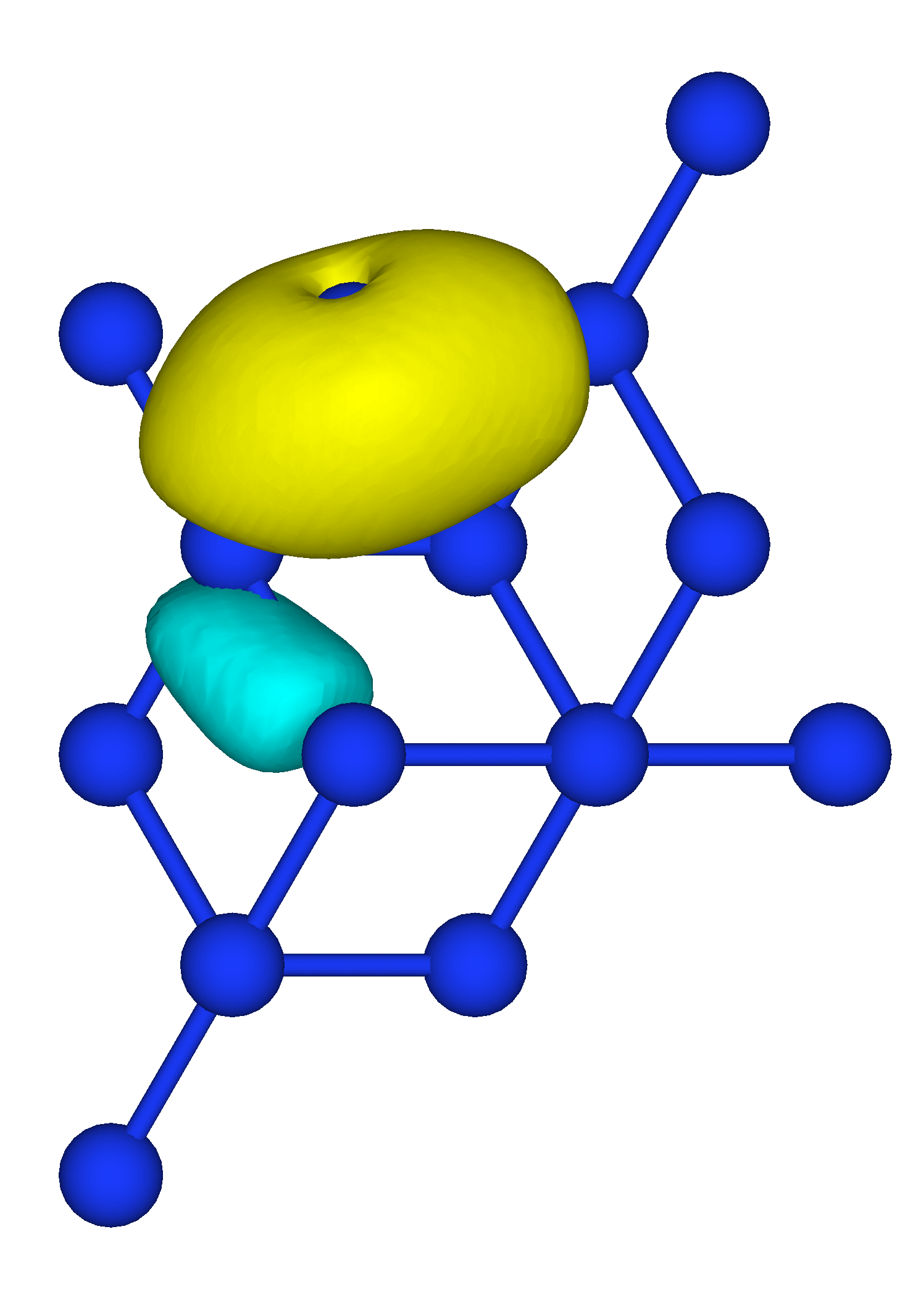}
  \caption{Column interpolation method}
  \label{column_silicon}
 \end{subfigure}
 \caption{One of the four Wannier functions of silicon, isosurface plot at $20\%$ of maximal value.}
 \label{WFsilicon}
 \end{figure}

In Table \ref{locSilicon}, we can see that the value of the localisation functional $\Omega$ is better for the logarithm method than for ours, but, after optimisation of the Marzari-Vanderbilt procedure~\cite{marzari2012maximally}, both methods agree.  

In Figure \ref{WFsilicon}, we display some Wannier functions computed
by both methods, before optimisation. The representation was done
through Wannier90 \cite{Wannier90} and VESTA \cite{VESTA}. The
localisation of both is not optimal, which is expected, but the
Wannier functions are still localised, and physically relevant.

\section*{Conclusion}

We presented a new method to construct localised Wannier functions. It
is proven to work even in the case of topological insulators which
causes most published algorithms to fail. In the ``easy'' cases, it
works similarly to the method of \cite{cances2017robust}. As that
method, it only localises Wannier functions across unit cells, and
does not attempt to localise the functions inside the unit cell. This
is problematic in the case of large unit cells, which is the case of
many real topological insulators. The efficient numerical construction
of Wannier functions in these cases remains therefore an open problem.

\bibliographystyle{alpha}
\bibliography{Wannier_Functions}

{\footnotesize

\begin{tabular}{rl}
(D. Gontier) & \textsc{Universit\'e Paris-Dauphine, PSL Research University, CEREMADE} \\
& 75775 Paris, France \\
& \textsl{E-mail address}: \href{mailto:gontier@ceremade.dauphine.fr}{\texttt{gontier@ceremade.dauphine.fr}} \\
\\
  (A. Levitt) & \textsc{Inria Paris and Universit\'e Paris-Est, CERMICS (ENPC)} \\
 &  F-75589 Paris Cedex 12, France\\
 &  \textsl{E-mail address}: \href{mailto:antoine.levitt@inria.fr}{\texttt{antoine.levitt@inria.fr}} \\
  \\
(S. Siraj-Dine) & \textsc{Universit\'e Paris-Est, CERMICS (ENPC) and
                 Inria Paris} \\
 &  F-77455 Marne-la-Vall\'ee, France\\
 &  \textsl{E-mail address}: \href{mailto:sami.siraj-dine@enpc.fr}{\texttt{sami.siraj-dine@enpc.fr}} \\
 \\
\end{tabular}

The PhD of Sami Siraj-Dine is supported by the Labex Bézout.
}
\end{document}